# Stabilities of generalized entropies

Sumiyoshi Abe[1], G. Kaniadakis[2], and A. M. Scarfone[2]

[1]*Institute of Physics, University of Tsukuba, Ibaraki 305-8571, Japan*

[2]*Dipartimento di Fìsica and Istituto Nazionale di Fìsica della Materia (INFM),*

*Politecnico di Torino, Corso Duca degli Abruzzi 24,*

*I-10129 Torino, Italy*

The generalized entropic measure, which is optimized by a given arbitrary distribution under the constraints on normalization of the distribution and the finite ordinary expectation value of a physical random quantity, is considered and its Lesche stability property (that is different from thermodynamic stability) is examined. A general condition, under which the generalized entropy becomes stable, is derived. Examples known in the literature, including the entropy for the stretched-exponential distribution, the quantum-group entropy, and the $\kappa$-entropy are discussed.

PACS numbers: 65.40.Gr, 02.50.-r, 05.20.-y, 05.90.+m

# I. INTRODUCTION

There is great diversity in statistical distributions observed in nature. This is apparently a challenge for traditional statistical mechanics. In view of traditional statistical mechanics based on Boltzmann-Gibbs-Shannon entropy, a significant number of distributions observed in complex systems are actually anomalous. Examples include granular materials, glassy systems, self-gravitating systems, biological systems, and seismicity. An important point here is that these anomalous distributions can persist for very long periods of time, much longer than typical time scales of underlying microscopic dynamics. This fact naturally leads to a question if there would be a framework for understanding such diverse statistical phenomena in a unified manner. In this respect, the principle of maximum entropy pioneered by Gibbs and Jaynes may be though of a one such [1]. Then, if one wishes to describe such anomalous distributions based on the principle of maximum entropy, there seem to be only two ways to be addressed. One is to modify the form of the constraints, and the other is to generalize the Boltzmann-Gibbs-Shannon entropy. The latter is the standing point that we take in the present work.

In a recent paper [2], one of the present authors has presented an entropy generating algorithm and has constructed a very general class of entropic measures which are optimized by given distributions under the appropriate constraints on normalization and the ordinary expectation value of a physical random variable such as the energy. In spite of its mathematical consistency of the discussion, however, it is still not clear if the entire class of such generalized entropies may behave as good measures.

In this paper, we examine the concept of stability proposed by Lesche in Ref. [3] (see also Ref. [4]), which should be satisfied by any physical entropic measure. We shall derive a general condition, under which the generalized entropy can satisfy the Lesche stability property (that is different from thermodynamic stability). The stability properties of a number of examples including the entropy for the stretched-exponential distributions [5], the quantum-group entropy [6], and the $\kappa$-entropy [7-10] are also discussed.

## II.   GENERALIZED ENTROPY

In Ref. [2], an algorithm has been presented for generating a generalized entropy which is optimized by a given arbitrary distribution under the constraints on normalization of the distribution and the ordinary expectation value of a physical quantity, $\{Q_i\}_{i=1, 2, \cdots, W}$, of interest (e.g., the internal energy), where $W$ is the number of microscopically accessible states.

Given a normalized distribution $\{p_i = f(\alpha + \beta Q_i)\}_{i=1, 2, \cdots, W}$, the corresponding generalized entropy optimized by it is constructed as follows:

$$S[p] = \int_{t_{min}}^{t_{max}} dt \ (1 - A[p;t]) + c. \tag{1}$$

Here, $\alpha$ and $\beta$ are the Lagrange multipliers associated with the constraints on normalization and the ordinary expectation value of $\{Q_i\}_{i=1, 2, \cdots, W}$, respectively. $A[p; t)$ is a quantity given by

$$A[p;t] = \sum_{i=1}^{W} \left(p_i - f(t)\right)_+ \tag{2}$$

with the notation

$$(x)_+ = \max\{0, x\}. \tag{3}$$

$c$ is the constant which should be determined in such a way that $S[p]$ vanishes for the completely ordered state, $p_i = p_i^{(0)} = \delta_{ij}$ ($1 \le j \le W$). $f(t)$ is a function that determines the form of the distribution, $p_i$. For the sake of simplicity, this function is assumed to be a monotonically-decreasing function with the range $(0, 1)$ and to satisfy the condition

$$\int_{t_{min}}^{t_{max}} dt\, f(t) < \infty, \tag{4}$$

where $(t_{min}, t_{max})$ is the domain of $f(t)$, i.e., $f(t) \to 1\, (0)$ as $t \to t_{min}\, (t_{max})$.

It can be seen that Eq. (1) is written in the following form:

$$S[p] = \sum_{i=1}^{W} \left[ p_i f^{-1}(p_i) - \int_{f^{-1}(0)}^{f^{-1}(p_i)} dt\, f(t) \right] + \int_{f^{-1}(0)}^{f^{-1}(1)} dt\, f(t) - f^{-1}(1), \tag{5}$$

where $f^{-1}$ is the inverse function of $f$. Moreover, if $f^{-1}$ is piecewise differentiable, as assumed here and hereafter, then $S[p]$ can be further rewritten as follows [11]:

$$S[p] = \sum_{i=1}^{W} \int_0^{p_i} dt\, f^{-1}(t) - \int_0^1 dt\, f^{-1}(t). \tag{6}$$

With this form, it is now evident that the stationarity condition on the functional, i.e., $\delta\left(S[p] - \alpha \sum_{i=1}^{W} p_i - \beta \sum_{i=1}^{W} p_i Q_i\right) = 0$, in fact yields the optimal distribution $\{p_i = f(\alpha + \beta Q_i)\}_{i=1,2,\cdots,W}$.

The construction in Eq. (1) with Eq. (2) is a general mathematical procedure for generating a concave functional of $\{p_i\}_{i=1,2,\cdots,W}$. In addition, as shown in Ref. [2], $S[p]$ satisfies the H-theorem for the master equation combined with the principle of microscopic reversibility.

## III. STABILITY CRITERION

It is not expected that whole class of generalized entropic measures expressed in the form in Eq. (1) or Eq. (6) are physically relevant, even though they are concave and satisfy the H-theorem. In order for an entropic measure to be experimentally robust, it is necessary for the measure to satisfy the stability condition proposed in Ref. [3]. This concept is stated as follows. Usually, what is experimentally measured is not directly a statistical entropy, $\Sigma$, itself but a distribution. Repeating the same experiment, an experimentalist will obtain a distribution, which may be slightly different from the previously obtained one. If $\Sigma$ is of physical relevance, then at least its value should not change drastically for two slightly different distributions, $\{p_i\}_{i=1,2,\cdots,W}$ and $\{p'_i\}_{w=1,2,\cdots,W}$. Mathematically, this implies

$$(\forall \varepsilon > 0) \quad (\exists \delta > 0) \quad \left( \| p - p' \|_1 < \delta \Rightarrow \left| \frac{\Sigma[p] - \Sigma[p']}{\Sigma_{\max}} \right| < \varepsilon \right) \tag{7}$$

for any value of W, where $\|A\|_1 = \sum_{i=1}^{W} |A_i|$ and $\Sigma_{\max}$ is the maximum value of $\Sigma$. To examine this condition for the quantity in Eq. (1), we analyze the following inequality:

$$|S[p] - S[p']| \leq \int_{t_{\min}}^{\tau} dt \, |A[p;t) - A[p';t)| + \int_{\tau}^{t_{\max}} dt \, |A[p;t) - A[p';t)|, \tag{8}$$

where $\tau$ satisfies

$$t_{\min} < f^{-1}(1/W) \leq \tau < t_{\max}. \tag{9}$$

To evaluate the right-hand side of Eq. (8), we notice the following properties:

$$|A[p;t) - A[p';t)| \leq \|p - p'\|_1, \tag{10}$$

$$|A[p;t) - A[p';t)| \leq W f(t) \qquad (t \geq f^{-1}(1/W)). \tag{11}$$

Using Eqs. (10) and (11) in Eq. (8), we find

$$|S[p] - S[p']| \leq G(\tau), \tag{12}$$

$$G(\tau) = (\tau - t_{min})\|p - p'\|_1 + W \int_{\tau}^{t_{max}} dt\, f(t). \tag{13}$$

Eq. (12) holds for any value of $\tau$ satisfying Eq. (9). Let us take $\tau = \tau_0$ which makes $G(\tau)$ minimum:

$$\tau_0 = f^{-1}\left(\|p - p'\|_1 / W\right). \tag{14}$$

Therefore, we have

$$|S[p] - S[p']| \leq \left\{ f^{-1}\left(\|p - p'\|_1 / W\right) - t_{min} \right\} \|p - p'\|_1$$

$$+ W \int_{f^{-1}(\|p-p'\|_1/W)}^{t_{max}} dt\, f(t)$$

$$= W t_{max} f(t_{max}) - t_{min} \|p - p'\|_1 - W \int_{\|p-p'\|_1/W}^{f(t_{max})} dt\, f^{-1}(t). \tag{15}$$

Noticing that Eq. (6) takes the following maximum value for the equiprobability

$$S_{max} = W \int_0^{1/W} dt\, f^{-1}(t) - \int_0^1 dt\, f^{-1}(t), \tag{16}$$

we have

$$\left|\frac{S[p]-S[p']}{S_{max}}\right| \leq B(\|p-p'\|_1, W), \tag{17}$$

$$B(\|p-p'\|_1, W) \equiv \frac{\int_0^{\|p-p'\|_1/W} dt\, f^{-1}(t) - (t_{min}/W)\|p-p'\|_1}{\int_0^{1/W} dt\, f^{-1}(t) - (1/W)\int_0^1 dt\, f^{-1}(t)}, \tag{18}$$

where we have used the fact that $t f(t)$ tends to vanish in the limit $t \to t_{max}$, due to Eq. (4) as well as the property, $f(t) \to 0$ $(t \to t_{max})$. Therefore, we conclude that the generalized entropy is stable in the thermodynamic limit, $W \to \infty$, if

$$\lim_{\|p-p'\|_1 \to +0} \lim_{W \to \infty} B(\|p-p'\|_1, W) = 0. \tag{19}$$

This is our main result. Notice that this order of taking the limits is essential for the Lesche stability criterion.

Notice that $B(\|p-p'\|_1, W)$ in Eq. (18) is an indeterminate form in the limit $W \to \infty$. A particular case when application of L'Hopital's rule once to this limit is sufficient, then we have

$$\left|\frac{S[p]-S[p']}{S_{max}}\right| \leq C\|p-p'\|_1, \tag{20}$$

$$C = \frac{f^{-1}(+0) - f^{-1}(1-0)}{f^{-1}(+0) - \int_0^1 dt\, f^{-1}(t)} . \tag{21}$$

So, in this case, taking $\delta$ as $\delta = \varepsilon / C$, we see that the generalized entropy satisfies the stability condition in Eq. (7).

Closing this section, we notice that in Ref. [12] the continuity and stability properties of a class of generalized entropies are discussed by employing an approach different from the present one.

### IV. EXAMPLES

In this section, we discuss some examples of stable generalized entropies known in the literature.

#### 1. Entropy for stretched-exponential distribution

In this case, $f(t)$ is taken to be

$$f(t) = \exp(-t^\gamma), \tag{22}$$

where $\gamma \in (0, 1)$ and $t \in (t_{\min}, t_{\max}) = (0, \infty)$. Substitution of this function into Eq. (6) gives rise to the following generalized entropy [2,5]:

$$S_{SE}[p] = \sum_{i=1}^{W} \Gamma(1+1/\gamma, -\ln p_i) - \Gamma(1+1/\gamma), \tag{23}$$

where $\Gamma(u, x)$ is for the incomplete gamma function of the second kind, $\Gamma(u, x) = \int_{x}^{\infty} dt\, t^{u-1} e^{-t}$, and $\Gamma(u) = \Gamma(u, 0)$ is the ordinary gamma function. Since $f^{-1}(t) = (-\ln t)^{1/\gamma}$ with $t \in (0, 1)$, $C$ in Eq. (21) is calculated to be

$$C = \lim_{t \to +0} \frac{f^{-1}(t)}{f^{-1}(t) - \Gamma(1+1/\gamma)}$$

$$= 1. \tag{24}$$

Therefore, taking $\delta = \varepsilon$, the entropy for the stretched-exponential distribution is seen to satisfy the Lesche stability condition.

In the particular case when $\gamma \to 1-0$, $S_{SE}[p]$ converges to the Boltzmann-Gibbs-Shannon entropy, $S_{BGS}[p] = -\sum_{i=1}^{W} p_i \ln p_i$, as it should do. Thus, as a byproduct, stability of the Boltzmann-Gibbs-Shannon entropy shown in Ref. [3] is ascertained.

2.  **Quantum-group entropy**

The quantum-group entropy is given by

$$S_{QG}[p] = -\sum_{i=1}^{W} \frac{(p_i)^q - (p_i)^{q^{-1}}}{q - q^{-1}}. \tag{25}$$

This quantity has been introduced in Ref. [6] and has been applied to generalized statistical-mechanical study of $q$-deformed oscillators. In this expression, $q$ is assumed to be positive. Since $S_{QG}[p]$ is symmetric under interchange, $q \leftrightarrow q^{-1}$, the range of $q$ can be reduced to $(0, 1)$. This quantity also converges to the Boltzmann-Gibbs Shannon entropy in the limit $q \to 1$.

The function, $f(t)$, defined on $(t_{min}, t_{max}) = (-1, \infty)$ associated with the quantum-group entropy is implicitly given as the inverse function of

$$f^{-1}(t) = -\frac{qt^{q-1} - q^{-1}t^{q^{-1}-1}}{q - q^{-1}}. \tag{26}$$

For this function, $C$ in Eq. (21) is still an indeterminate form. Accordingly $B(\|p - p'\|_1, W)$ in Eq. (18) is evaluated directly as follows:

$$B(\|p - p'\|_1, W) = \frac{\int_0^{\|p-p'\|_1/W} dt\, f^{-1}(t) + (1/W)\|p - p'\|_1}{\int_0^{1/W} dt\, f^{-1}(t)}$$

$$= \frac{(\|p - p'\|_1/W)^q - (\|p - p'\|_1/W)^{q^{-1}} - (q - q^{-1})(1/W)\|p - p'\|_1}{W^{-q} - W^{-q^{-1}}}$$

$$\to (\|p - p'\|_1)^q \qquad (W \to \infty). \tag{27}$$

Therefore, taking $\delta = \varepsilon^{1/q}$, we see that the quantum-group entropy is stable.

**3.  $\kappa$-entropy**

The $\kappa$-entropy has been introduced in Ref. [7] and has been applied to systems described by statistical distributions having a power-law asymptotic behavior. In Ref. [7], the following one-parameter generalizations of the ordinary exponential and logarithmic functions have been proposed:

$$\exp_{\{\kappa\}}(t) = \left(\sqrt{1+\kappa^2 t^2} + \kappa t\right)^{1/\kappa}, \tag{28}$$

$$\ln_{\{\kappa\}}(t) = \frac{t^\kappa - t^{-\kappa}}{2\kappa}, \tag{29}$$

from which the ordinary exponential and logarithmic functions are respectively reproduced in the limit $\kappa \to +0$. $\kappa$ should be in the range $(-1, 1)$. On the other hand, $\exp_{\{\kappa\}}(t)$ is defined for $t \in (-\infty, \infty)$ and $\ln_{\{\kappa\}}(t)$ for $t \in (0, \infty)$. Both of these functions are symmetric under interchange $\kappa \leftrightarrow -\kappa$. [For $\exp_{\{\kappa\}}(t)$, this interchange is combined with the reversal of *t*.] Therefore, the range of $\kappa$ can be taken to be $(0, 1)$.

It is a simple task to verify that if we choose

$$f(t) = \exp_{\{\kappa\}}(-t), \tag{30}$$

$$f^{-1}(t) = -\ln_{\{\kappa\}}(t) \qquad \text{for } t \in (t_{\min}, t_{\max}) = (0, \infty) \tag{31}$$

then we obtain the $\kappa$-entropy given by

$$S_\kappa[p] = -\sum_{i=1}^{W} \left\{ c_{-\kappa} \left[(p_i)^{1-\kappa} - p_i\right] + c_\kappa \left[(p_i)^{1+\kappa} - p_i\right]\right\}, \tag{32}$$

$$c_\kappa = \frac{1}{2}\left(\frac{1}{\kappa} + \frac{1}{1+\kappa}\right). \tag{33}$$

Substituting Eq. (31) into Eq. (18), we find

$$B(\|p - p'\|_1, W) \to \left(\|p - p'\|_1\right)^{1-\kappa} \qquad (W \to \infty). \tag{33}$$

Therefore, setting $\delta = \varepsilon^{1/(1-\kappa)}$, we conclude that the $\kappa$-entropy is stable [10]. It is worth mentioning that the $\kappa$-entropy becomes reduced to the Boltzmann-Gibbs-Shannon entropy in the limit $\kappa \to +0$.

## V.  CONCLUDING REMARKS

We have discussed the generalized entropy optimized by a given arbitrary distribution under the constraints on normalization of the distribution and the ordinary expectation value of a physical random quantity. We have examined its Lesche stability property and have derived a general condition, under which the generalized entropy becomes stable.

We have also discussed some examples of entropic measures known in the literature and have shown their stabilities.

In recent years, much attention has been paid to the Tsallis entropy [13], which has been employed for generalizing Boltzmann-Gibbs statistical mechanics for nonextensive systems [14-16]. It has been shown in Ref. [4] that the Tsallis entropy is stable.

There are also unstable entropies. Examples are the Rényi entropy [17] and the so-called normalized Tsallis entropy [18,19], whose instabilities have been shown in Ref. [3] and Ref. [4], respectively. Quite interestingly, these quantities cannot be expressed in the form in Eq. (1) and are not concave if their entropic indices are larger than unity.

Finally, we point out that mathematically the Lesche stability property is equivalent to uniform continuity of the functional under consideration. The problem of continuity itself has recently been studied in Refs. [20,21], where the Boltzmann-Gibbs-Shannon entropy is shown not to be continuous for infinite microscopic states.

**Acknowledgment**

S.A. was supported in part by the Grant-in-Aid for Scientific Research of Japan Society for the Promotion of Science.